# Combining predictors of natively unfolded proteins to detect a twilight zone between order and disorder in generic datasets.

Antonio Deiana[1], Andrea Giansanti[1]*

[1]Department of Physics, La Sapienza, University of Rome, P.le A. Moro 5, 00185, Rome, Italy

*Corresponding author

Email addresses:
    AD: Antonio.Deiana@roma1.infn.it
    AG: Andrea.Giansanti@roma1.infn.it

## Abstract

Natively unfolded proteins lack a well defined three dimensional structure but have important biological functions, suggesting a re-assignment of the structure-function paradigm. Many proteins have amino acidic compositions compatible both with the folded and unfolded status, and belong to a twilight zone between order and disorder. This makes difficult a dichotomic classification of protein sequences into folded and natively unfolded ones. In this methodological paper dichotomic folding indexes are considered: hydrophobicity-charge, mean packing, mean pairwise energy, Poodle-W and a new global index, that is called here *gVSL2*, based on the local disorder predictor VSL2. The performance of these indexes is evaluated on different datasets. Poodle-W, *gVSL2* and mean pairwise energy have good performance and stability in all the datasets considered and are combined into a strictly unanimous combination score $S_{SU}$, that leaves proteins unclassified when the consensus of all combined indexes is not reached. The unclassified proteins: i) belong to an overlap region in the vector space of amino acidic compositions occupied by both folded and unfolded proteins; ii) are composed by approximately the same number of order-promoting and disorder-promoting amino acids; iii) have a mean flexibility intermediate between that of folded and that of unfolded proteins. These proteins reasonably have physical properties intermediate between those of folded and those of natively unfolded proteins and their structural properties and evolutionary history are worth to be investigated.

## Introduction

For long time it has been thought that the existence of a stable three dimensional structure is necessary for a protein molecule to be functional. Nevertheless, the evidence that many proteins are unfolded in their functional state has induced, in the last decade, a re-assignment of the structure-function paradigm [1]. Natively unfolded proteins, also known as "intrinsically unstructured" or "intrinsically disordered" [1-3] lack a well defined tertiary structure, being functional in states made of an ensemble of flexible conformations. It is known that these proteins are involved in important cellular processes like signalling, targeting and DNA binding [1-8]. It has been suggested that they may play critical roles in the development of cancer [9, 10] and in some amyloidotic diseases [10-12]; moreover, the absence of a rigid structure allows them to bind different targets with high specificity and low affinity, suggesting that they may be hubs in protein interaction networks [13-16]. The current release 4.9 of the Disprot database (http://www.disprot.org/; see also [17, 18]) contains 523 proteins with different "flavors" of disorder [19].

A great effort has been made to predict natively unfolded proteins through an *ab-initio* analysis of their amino acidic sequences. Many methods have been proposed (for reviews see [20, 21]), both to predict disordered amino acids of a protein, i.e. amino acids whose atoms are hard to localize by X-ray crystallography or NMR spectroscopy [22, 23], and to define synthetic binary scalar indexes that express the probability that a protein has a global tendency to fold or to remain unfolded. It has been observed that predictors of natively unfolded proteins are, basically, functionals of the amino acidic composition [24]. Szilagyi *et al.* reported that hydrophobicity and charge distributions of folded and natively unfolded proteins overlap, and, on a hydrophobicity/charge plane, there is an area occupied by proteins from both groups [24]. This suggests that, in the vector space of amino acidic compositions, it is not possible to find a hyper-plane separating the points corresponding to sequences of folded proteins and those corresponding to sequences of unfolded proteins, but there exists an overlap volume, that is reasonable to identify with a "twilight zone" between order and disorder [24, 25].

In this work we are interested in finding an operational method to identify proteins in this twilight zone.

Uversky *et al.* propose to analyse the mean hydrophobicity and mean net charge of the protein sequences [26]; following their method, Prilusky *et al.* developed a web-based server named FoldIndex [27]. Galzitskaya *et al.* propose to use the mean packing of a protein sequence as an indicator of its folding status [28-30]. Recently, mean pairwise energy was used to effectively discriminate folded proteins from natively unfolded ones in a peculiar set of 39 protein complexes [31, 32]. A more refined approach is Poodle-W by Shimizu *et al.* [33]; they analyse the amino acid composition of the protein sequences with a spectral graph transducer [34].

We compared the performances of these predictors on several datasets. We also defined a global binary folding index, named here *gVSL2*, using VSL2 [35, 36], a predictor of disordered amino acids that excellently performed in the CASP7 experiment [37] and which is an evolution of the predictors that evidenced different flavours of disorder [19].

We observed that in many cases a number of sequences were differently classified by different indexes. To identify these proteins we introduced, by combining several indexes, a strictly unanimous consensus score $S_{SU}$ that leaves unclassified a protein if at least two indexes disagree in classifying it. We verified that proteins unclassified by $S_{SU}$ span, on a hydrophobicity/charge plane, an overlap area occupied both by folded and natively unfolded proteins (figures 2 and 3). We checked also that these proteins have amino acidic frequencies intermediate between those of folded and those of natively unfolded proteins (figures 4, 5), so we have concluded that they belong to a twilight zone in the space of amino acidic composition. Proteins in this twilight zone have a flexibility intermediate between that of folded and that of natively unfolded proteins and have a dependence of chain length similar to that reported by Szilagyi *et al.* [24] (figures 7 and 8).

We used $S_{SU}$ to scan several genomes from Archaea, Bacteria and Eukarya looking for natively unfolded proteins; the obtained percentages are similar to those already reported in the literature [38, 39]; through $S_{SU}$ we found robust scaling laws both in the classified and unclassified sets of proteins, of possible significance for studies in molecular evolution [40, 41].

## Methods

### Datasets

Set A was selected by Prilusky *et al.* to test FoldIndex [27]. It is composed by 151 folded proteins and 39 natively unfolded proteins. Folded proteins are extracted from OCA Protein Data Bank (http://bioportal.weizmann.ac.il/oca), selecting those with less than 10% of sequence identity and lengths comprised between 50 and 200 residues. Proteins with disulfide bridges, hetero-groups or other non-protein elements are excluded; moreover, only X-ray structures with a percentage of disordered amino acids below 5% are considered. Unfolded proteins are selected by searching the literature for experimentally certified cases. Set B was selected by Shimizu *et al.* to train and test Poodle-W [33]. It is composed by 526 folded proteins selected from the PDB, and 81 natively unfolded proteins selected from the DisProt database [17, 18].

Folded proteins are extracted from the PDB, among those with: i) resolution better than 2 Å; ii) R-factor below 20% and refined through Refmac5, SHELXL97 or CNS; iii) a percentage of disorder below 5% and no disordered amino acids in the central area ( i.e. between the 10th residue from the N-terminal and the 10th residue from the C-terminal). Natively unfolded proteins are extracted from DisProt version 3.3 among those with a percentage of disordered amino acids above 70%. Both folded proteins and natively unfolded ones have a sequence identity below 30%, determined using BLASTClust.

Set C was selected for the present work. Folded proteins are extracted from PDBSelect25 [42, 43], version October 2007, which contains 3694 proteins with sequence identity lower than 25%. Structures with a resolution above 2 Å and an R-factor above 20% are excluded. A restricted list of 2369 is obtained, 1573 of which with a percentage of disordered amino acids below 5%. Operationally, a residue is considered as disordered if it is present in the SEQRES but not in the ATOM field of the PDB file [37, 38]. A list of 81 natively unfolded proteins, with at least 70% of disordered amino acids and sequence identity below 25%, are extracted from the DisProt database [17, 18], version 3.6. Note that the 81 unfolded proteins in this set, coincide only in part with the 81 unfolded proteins selected, with a more refined procedure, in [33]. The list of PDB and DisProt entries of the proteins in set C are available as supplementary material.

### HQ index

The method by Uversky *et al.* [26], revisited by Prilusky *et* al. [27] is followed here. From mean hydrophobicity $<H>$ and mean net charge $<Q>$ of the protein sequences, the following index is evaluated:

$$HQ = 2.785\ <H> - |\ <Q>\ | - 1.151$$

A protein is considered as natively unfolded if *HQ* is negative or zero, otherwise it is considered as folded.

### Mean packing

The mean packing of a protein sequence is the arithmetic mean of the packing values of each amino acid. The packing index has been introduced by Galzitskaya *et al.* [29], it is evaluated by counting, for each type of residue, the residues located within a distance of 8 Å, and averaging over a large dataset of structures. The packing index is here assigned to the central residue as the average over a sliding window of length 11. A protein is considered here as natively unfolded if its mean packing is below 20.55, otherwise it is considered as folded. This threshold is optimized so to obtain a sensitivity of at least 0.80 and a level of false positives as low as possible. We did not observe a significant improvement of the performance changing the length of the sliding window.

### Mean pairwise energy

The method by Dosztanyi *et al.* [31] is followed, as implemented in IUPred [44]. The pairwise energy of an amino acid expresses its "contact interaction" with the amino acids located, downward and upward, from 2 to 100 position apart, along a given sequence. The pairwise energy of amino acid *i* at position *p* is given by:

$$e_i^{(p)} = \sum_{j=1}^{20} P_{ij} n_j^{(p)}$$

where $n_j^{(p)}$ is the frequency of amino acid *j* in a window of length up to 100 around position *p*, taking into account the limitations on both sides, due to the length of the protein. The generic element $P_{ij}$ of the "energy predictor matrix" *P* expresses the expected contact interaction energy between amino acid *i* and *j*. Pairwise energy values are averaged over a window of 21 amino acids and the average is assigned to the central residue at position *p* in the sequence. Finally, the arithmetic mean of the pairwise energy values of all the amino acids gives the global mean pairwise energy of the protein.

A protein is considered here as natively unfolded if its mean pairwise energy is above -0.37 a.e.u., otherwise it is considered as folded. This threshold is optimized so to obtain a sensitivity of at least 0.80 and a level of false positives as low as possible.

**Index *gVSL2* derived from VSL2**

VSL2 [35, 36] is a disorder predictor that assigns to each amino acid of a protein its propensity to be disordered, estimated using a combination of support vector machines. The score from VSL2 is normalized between 0 and 1 and an amino acid is considered disordered if its value is above 0.5.

The global index *gVSL2* is evaluated by assigning to the central residue the average of VSL2 scores (variant VSL2B) over sliding windows of 11 residues, and then taking the average value over all the residues. A protein is classified here as natively unfolded if its *gVSL2* value is above 0.5.

**Poodle-W**

Poodle-W [33] is a predictor of natively unfolded proteins based on a Spectral Graph Transducer semi-supervised learning machine [34]. It analyses the amino acid compositions of both proteins with known structural properties and with unknown structural properties, and it constructs a k-nearest neighbour graph with sequences as vertices and similarity among sequences as edges; then it cuts the graph into two sub-graphs so as to minimize the misclassification of the sequences with known structure and the sum of the edges weights of the graph; the resulting sub-graphs identify the predicted folded and natively unfolded proteins.

**Statistics for the amino acid composition of protein sequences**

The approach of Romero *et al.* [45] is followed. In a dataset containing $N_{sequ}$ sequences, let $n_j$ be the number of the amino acids of the j-th sequence and $n_{ij}$ the number of occurrence of the i-th amino acid in the j-th sequence. The frequency of the i-th amino acid in the j-th sequence is given by:

$$f_{ij} = \frac{n_{ij}}{n_j}$$

with variance

$$\mathrm{var}(f_{ij}) = \frac{f_{ij} \cdot (1 - f_{ij})}{n_j}$$

The frequency of the i-th amino acid in the dataset analysed is computed as:

$$f_i = \frac{\sum_{j=1}^{N_{sequ}} n_j \cdot f_{ij}}{\sum_{j=1}^{N_{sequ}} n_j} = \frac{\sum_{j=1}^{N_{sequ}} n_j \cdot f_{ij}}{N_{aa}}$$

where $N_{aa}$ is the number of the amino acids in the dataset. The corresponding variance is given by:

$$\mathrm{var}(f_i) = \frac{\sum_{j=1}^{N_{sequ}} n_j^2 \, \mathrm{var}(f_{ij})}{\left( \sum_{j=1}^{N_{sequ}} n_j \right)^2}$$

**Log-odds ratio of the likelihoods that a sequence has amino acidic composition typical of folded and unfolded proteins**

To compute this score, one usually refers to a probabilistic model [46]. One assumes to have reliable estimates of the probability of occurrence of each amino acid $a$ in folded and unfolded proteins $\{\pi_a^{(F)}\}$ and $\{\pi_a^{(U)}\}$, respectively. A folded protein can be thought of as if its sequence is sampled from $\{\pi_a^{(F)}\}$ through a sequence of independent extractions. The likelihood that a sequence has amino acidic composition typical of a folded protein is:

$$L_F = \prod_{a=1}^{20} \left( \pi_a^{(F)} \right)^{n_a}$$

where $\pi_a^{(F)}$ is the probability of amino acid $a$ (estimated from the occurrences of $a$ in a convenient sample of folded proteins) and $n_a$ is the occurrence of amino acid $a$ in the sequence. Similarly we can define $L_U$.
The log-odds ratio of a given sequence is then:

$$S = \ln \frac{L_F}{L_U} = \sum_{a=1}^{20} n_a \ln \frac{\pi_a^{(F)}}{\pi_a^{(U)}}$$

Order-promoting amino (disorder-promoting) acids contribute with positive (negative) terms to S, since their ratios $\pi_a^{(F)}/\pi_a^{(U)}$ are bigger (lower) than one. Therefore, S is positive (negative) if the protein is composed mainly by order-promoting (disorder-promoting) amino acids. When a protein is composed by approximately the same number of order-promoting and disorder-promoting amino acids, its S score has a value close to zero.

**Mean flexibility**

The mean flexibility of a protein is the arithmetic mean of the flexibility values of its amino acids. We use here the flexibility scale by Smith *et al.* [47]; they derive their flexibility scale through an analysis of B-factors of 290 proteins selected from PDB Select 25 version August 1998. The flexibility index is here assigned to the central residue as the average over a sliding window of length 11. We did not observe a significant difference in the behaviour of figure 7 changing the length of the sliding window.

**Indicators of performance**

The performance of the predictors is evaluated using very common indicators [37, 48]:

- Sensitivity: $S_n = \dfrac{TP}{TP+FN}$
- Specificity: $S_p = \dfrac{TN}{TN+FP}$
- False predictions: $f_p = 1 - S_p = \dfrac{FP}{TN+FP}$
- Area Under Curve (AUC) of a Receiver Operating Characteristic (ROC) curve

where TP stands for the true positives, TN the true negatives, FP the false positives and FN for the false negatives.

$S_n$ and $S_p$ express the fraction of correctly predicted proteins in a dataset; more precisely, $S_n$ is the fraction of correctly identified natively unfolded proteins, whereas $S_p$ is the percentage of correctly identified folded proteins. $f_p$ expresses the fraction of folded proteins that are classified by the index as natively unfolded. Receiver operating characteristic (ROC) curves are widely used to evaluate the performance of binary folding indexes and are obtained by plotting the fraction of true positive predictions versus the fraction of false negative predictions for each threshold value of the probability to be disordered. The performance of the predictor is evaluated through the area under the curve (AUC), generally computed by the trapezoid rule; the AUC is independent from discriminative thresholds. The AUC values were computed using ROCR [49].

# Results

**Performances of single indexes of folding**

To compare the performance of the synthetic predictors defined in the Methods section we computed them on set A, that Prilusky *et al.* selected to test FoldIndex [27]. The results are reported in table 1. From the AUC values shown in table 1 it is possible to rank the single indexes. We observe that mean pairwise energy, *gVSL2* and Poodle-W have about the same performance and are the best three.

To test the stability of these results over different datasets, we repeated the experiment using two other groups of proteins: set B, used by Shimizu *et al.* to test Poodle-W [33] and set C, our own selection of proteins. Set B contains 526 folded proteins with a level of disorder below 5% and 81 natively unfolded proteins with a level of disorder above 70%; it has been built as a very discriminative set, aiming at including either *fully ordered* or *fully disordered* proteins. Set C was compiled to test the ability of previously proposed methods, to separate, on one side, fully ordered folded proteins but also proteins containing a percentage of disordered amino acids above 5%, from, on the other side, fully disordered proteins containing more than 70% of disordered residues. It contains 2369 folded proteins, 1573 fully ordered with a level of disorder below 5% and the remaining 796 proteins with higher percentage of disordered amino acids. Set C contains also 81 natively unfolded sequences with a level of disorder above 70% (see Methods). The performances of the single folding indexes on set B and C are reported in tables 2 and 3, respectively. The AUC values confirm that Poodle-W, *gVSL2* and mean pairwise energy $<E_c>$ are the best performing on both sets. Note also that Poodle-W and $<E_c>$ have a low level of false positives $f_p$, whereas *gVSL2* tends to have a higher sensitivity $S_n$.

It is evident the lower performance of all the indexes on set C, with respect to set B. As said above, the protocol followed by Shimizu *et al*. [33] is more restrictive; their set is more *polarized*, containing folded proteins whose average composition is more different from that of unfolded than in set C (as shown in the hydrophobicity/charge plot of figure 3), which, being more generic, contains more untypical folded sequences. Poodle-W confirms itself as the relatively best single index, both on the optimized dataset B and on the more generic dataset C.

**A consensus score to detect a twilight zone of amino acidic composition**

In figure 1 we compare, in a Venn diagram, the results of the predictions obtained by mean pairwise energy $<E_c>$, *gVSL2* and Poodle-W for the 2450 proteins in set C. We see that the three indexes predict in the same folding class 1785 proteins, and that the indexes partly overlap, i.e. $<E_c>$ and *gVSL2* predict in the same folding class 231 proteins, and there are other 434 proteins on which at least two indexes of folding agree. On the other hand, there are proteins that are uniquely predicted by each single folding index: 224 by $<E_c>$, 210 by *gVSL2* and 231 by Poodle-W. So, globally, *there are 665 proteins on which at least two indexes disagree*.

It is reasonable to think that proteins differently classified by two single folding indexes have amino acidic frequencies different from those typical of folded and natively unfolded proteins, then they are natural candidates to belong to a twilight zone in the vector space of amino acidic compositions. To identify these proteins we used a consensus score. The use of combination scores is not new [39, 50] and a meta-server is also available [51]. We introduced a combination rule that we called *strictly unanimous score*, $S_{SU}$. We decided to combine mean pairwise energy, *gVSL2* and Poodle-W because, among the global indexes here considered, they are the best performing ones (see tables 1, 2 and 3). $S_{SU}$ classifies a protein as folded if *all* the indexes agree in predicting it as folded; conversely, it classifies a protein as natively unfolded if *all* the indexes agree in predicting it as natively unfolded; proteins are left *unclassified* when two indexes disagree. $S_{SU}$ classifies the majority of the sequences analysed (see *nc* in table 4): of the 2369 folded proteins, 646 are left unclassified, whereas of the 78 natively unfolded proteins, 19 are left unclassified; overall, the unclassified proteins are therefore 665, 27% of all proteins in set C. In set B the percentage of unclassified proteins is 11%.

Note that, due to the clause of unanimity, the performance of the single indexes on sets B and C purged from the unclassified proteins coincides with that of $S_{SU}$ and it is improved (compare tables 2 and 3 with table 4). This improvement after filtering can be obtained only if one combines in $S_{SU}$ a set of indexes of comparable performances, as discussed in the second subsection of the Discussion. We observe, in table 4, a relatively high sensitivity, and a low level of false positives, suggesting that removing the unclassified proteins is a filter of false predictions. We checked that 68% of the false predictions of $<E_c>$, 74% of those of *gVSL2* and 54% of those of Poodle-W are unclassified by $S_{SU}$. Note also the small fraction of unclassified unfolded proteins; that is essentially due to the fact that unfolded proteins are less numerous in the datasets.

Since unclassified proteins by $S_{SU}$ are often false predictions of the single indexes, they should have an untypical amino acidic composition. We checked that, in figures 2 and 3, using the hydrophobicity/charge plane, which is a projection of the space of amino acidic compositions. Figure 2 refers to set B, figure 3 to set C. Points in green and in black correspond to the proteins unclassified by $S_{SU}$. To give all the information we have distinguished with green points folded unclassified proteins, and with black points unfolded unclassified proteins. Note that in both figures (see insets) the centroids of the areas spanned by the three groups of proteins are well separated and aligned; those referring to unclassified proteins (green) are intermediate. Interestingly, in the more polarized set B, the proteins unclassified by $S_{SU}$ fall in a more restricted and well defined overlap area. These observations confirm that proteins we assign to the twilight zone belong to an overlap volume in the amino acidic composition space.

**Amino acidic composition of proteins in the twilight zone.**

It is interesting to check directly whether proteins which are unclassified by $S_{SU}$ have amino acidic frequencies intermediate between those of folded and those of natively unfolded proteins. We computed the frequencies of the amino acids for the three groups of proteins in set C. The results are reported in figure 4. We see that W, C, F, I, Y, V, L, H and N are more present in folded than in unfolded proteins, whereas A, R, Q, S, P, E, K are more frequent in unfolded than in folded proteins; this result is consistent with that reported in the literature [19, 45]. Following Romero *et al.* [45], we indicate the first group of amino acids as order-promoting and the second group as disorder-promoting. We observe also that the unclassified proteins by $S_{SU}$ have a peculiar composition with respect to folded and natively unfolded ones; more precisely, the frequencies of W, F, I, Y, V, L, Q, S, P, E and K are comprised between those of folded and natively unfolded proteins, whereas M, T, A, N and D have frequencies similar in the three groups of sequences. By looking at figure 4 one would conclude that the amino acidic composition of the unclassified proteins is substantially intermediate between those of folded and natively unfolded proteins. However, this could only signals that the unclassified proteins are a balanced mixture of folded and unfolded proteins. To exclude this hypothesis we checked that only a small fraction of unclassified proteins belong to the set of natively unfolded sequences (see previous section and table 4).

We further characterized the amino acidic composition of the proteins in set C by computing the log-odds ratio S of the likelihoods, for a sequence, of being composed mainly by order-promoting or disorder-promoting amino acids (see Methods for details). Positive (negative) values of S indicate that the sequence is composed mainly by order-promoting (disorder-promoting) amino acids [46]. The distributions of log-odds ratios are reported in figure 5 for sequences that are either classified by $S_{SU}$ or left unclassified. Note that, visually, the S scores are not normally distributed in each group of sequences; their statistics are different as checked by Wilcoxon's rank-sum

- 9 -

tests (p-values of the order of $10^{-16}$). As expected, proteins classified as folded have positive log-odds ratios (mean S equal to 6.98 ± 0.16, the median is 5.20). Proteins classified as unfolded have negative log-odds ratios (mean S is -6.87 ± 1.04 and the median is -3.28). Unclassified proteins have values of the log-odds ratio distributed around zero (mean 0.93 ± 0.16, median 0.52), between those of folded and natively unfolded proteins. This means that they are composed approximately by the same number of order-promoting and disorder-promoting amino acids, i.e. their amino acidic composition is intermediate between that of folded and natively unfolded proteins. Note that, in figure 4, the frequencies of C, H, R and G in the unclassified proteins, are not comprised between those in the other two classes but are even higher than those. This means that the amino acidic composition of the unclassified proteins is not strictly *in between* those of folded and natively unfolded proteins. The important point to make here is: the composition in the twilight zone is different, possibly corresponding to different folding propensities.

## Discussion

### General remarks

Most global predictors of disorder rely on the amino acid composition of protein sequences [24]. The distribution of the amino acids in folded proteins partly overlaps with that in natively unfolded proteins, moreover, as also noted in [24], the existence of two-state homodimers, that are disordered as monomers but fold on homo-dimerization, indicates that composition alone does not determine the folding status of a sequence. In this paper we have presented the unanimous consensus score $S_{SU}$; this score effectively selects proteins that, in the vector space of amino acidic composition, belong to an overlap volume occupied both by folded and natively unfolded proteins. This overlap volume has been identified and theoretically investigated by Szilagyi *et al.* [24] as a twilight zone between order and disorder. Proteins in the twilight zone, as identified by $S_{SU}$, have a distribution of the log-odds ratios S concentrated around zero, statistically different and intermediate between the S distributions of folded and unfolded proteins. This suggests that the proteins in the twilight zone are those with approximately the same number of order-promoting and disorder-promoting amino acids..

It is important to point out that $S_{SU}$ assigns proteins to the class of folded, unfolded or unclassified considering only their amino acidic composition and not the percentage of disordered amino acids they contain. So, proteins unclassified by $S_{SU}$ generally do not have a higher fraction of disordered amino acids than proteins classified as folded. In figure 6 we report the distribution of disordered amino acids in proteins assigned by $S_{SU}$ to the class of folded and unclassified proteins; it is evident that both classes have the same distribution. We checked that, in set C, of the 796 proteins with a percentage of disordered amino acids higher than 5%, 541 are classified by $S_{SU}$ as folded, 45 as unfolded and 210 are unclassified. We also checked that of the 294 complexed proteins alluded to in the results section and in the supplementary materials, 218 are classified as folded, 18 as unfolded and 58 are unclassified. So, proteins in the twilight zone cannot directly be identified with proteins with long flexible segments or loops. In figure 7 we report the distribution of mean flexibility in folded, unfolded and unclassified proteins (see Methods for definitions). Interestingly, unclassified proteins exhibit a mean flexibility (-0.465 ± 0.001) which is intermediate between that of folded (-0.486 ± 0.001) and natively unfolded proteins (-0.431 ± 0.002). This suggests that the mean flexibility of proteins in the twilight zone is not directly related to the number of disordered amino acids. It will be interesting, on a case-by-case study, to investigate how the distribution of disordered amino acids correlates with the localization of flexible tracts in the proteins of the twilight zone.

An interesting point is raised by Szilagyi et al. [24], who underscore the relevance of the chain length for the tendency of a protein to be folded or unfolded. This suggested us to look at the distribution of lengths in the proteins of the three classes: folded, unfolded and unclassified (twilight zone). It is interesting to note, in figure 8, that in all cases there is a scaling, power law, but the scaling exponents are quite different in the three classes; remarkably the twilight zone has the more negative scaling exponent (-3.3 ± 0.2), indicating that proteins in this class tend to be shorter than folded and unfolded ones. We believe that $S_{SU}$ can be useful as a starting point for further studies of the proteins lying in the region between order and disorder.

We already used $S_{SU}$ in a previous work [41], scanning several genomes from different kingdoms; we obtained percentages of natively unfolded proteins of about 0.8% for Archaea, 3.7% for Bacteria and 23.4% for Eukarya, consistent with those previously reported [38, 39]. The percentage of unclassified proteins is of 5.1% for Archaea, 7.4% for Bacteria and 15.8% for Eukarya. We also found scaling laws: the scaling exponents of the percentage of disordered proteins as a function of the number of proteins in the genome is 1.81 ± 0.10, whereas the percentage of unclassified proteins scales with an exponent 1.29 ± 0.05 [41]. In that analysis we combined mean packing, mean pairwise energy and $gVSL2$, but we did not use Poodle-W, since it was then unavailable to us. We are planning to extend that research using the operational combination of indexes proposed in the present study.

In the next section, at last, we propose a generic protocol on how to combine indexes into a strictly unanimous score.

**A protocol for the combination of indexes able to select a reliable subset of proteins in the twilight zone.**

The extent of the twilight zone made of the unclassified proteins depends on the particular choice of the combined indexes. We checked that if one combines indexes with relatively low performance with more performing ones then only a small fraction of unclassified proteins corresponds to the false predictions of most single indexes. In this case many mistakenly classified proteins by the poor performing indexes are just unrecognized true positives and true negatives of the best performing ones, and then the twilight zone is poorly characterized. To avoid that we propose the following simple protocol: the folding indexes are initially ordered by decreasing values of the AUC; then one or more indexes are subsequently used to form consensus scores, monitoring the parallel decrease of $f_p$. $S_{SU}^{(1)}$ is just the best performing single index; $S_{SU}^{(2)}$ is the combination of the first with the second best index; $S_{SU}^{(3)}$ the combination of the first three indexes, and so on. An enhancement of the performance of the indexes on the more and more polarized subsets is expected at each step, together with a parallel increase in the percentage of unclassified proteins. In fact, these trends are shown in table 5 for the indexes considered in this work; by looking at this table it is clear that, from step 3 on, $f_p$ tends to saturate, whereas the fraction of unclassified proteins increases. As a rule of thumb we propose to add indexes until the $f_p$ starts to saturate and this criterion we adopted in selecting Poodle-W, $gVSL2$ and mean pairwise energy $<E_c>$ in the present study.

# Conclusions

It has been pointed out that the separation between order and disorder in proteins is not sharp [24, 25], but there exists a twilight zone between them. In Szilagyi *et al.* [24] the twilight zone is identified as an overlap volume in the space of amino acidic compositions, occupied both by folded and natively unfolded proteins. In this work we used the consensus score $S_{SU}$ to select, operationally, proteins belonging to the twilight zone, out of a generic dataset. The method can be easily adopted for large database screening, since it is easy to implement and computationally efficient. Our results show that proteins unclassified by $S_{SU}$: i) belong to an overlap region in the vector space of amino acidic compositions occupied by both folded and unfolded proteins; ii) are composed by approximately the same number of order-promoting and disorder-promoting amino acids; iii) have a mean flexibility intermediate between that of folded and that of unfolded proteins. These last remarks indicate these proteins belong to a class that possibly has physical properties intermediate between those of folded and those of natively unfolded proteins.

# Acknowledgements

The authors gratefully thank Dr. K. Shimizu for her kindness in sending the list of proteins used to train and evaluate Poodle-W and for her collaboration in classifying for us, with Poodle-W, proteins in dataset C. The authors also thank Dr. Z. Dosztanyi for sending a draft of paper [21], and for providing the IUPred code.

# References

1. Wright P, Dyson HJ: **Intrinsically unstructured proteins: re-assessign the protein structure-function paradigm.** *J. Mol. Biol.* 1999, 293: 321-331.
2. Dyson HJ, Wright P: **Intrinsically unstructured proteins and their functions**. *Nat. Rev. Mol. Cell. Biol.* 2005, 6: 197-208
3. Dunker A, Lawson J, Brown C, Romero P, Oh J, Oldfield C, Campen A, Ratliffl C, Hipps K, Ausio J, Nissen M, Reeves R, Kang C, Kissinger C, Bailey R, Griswold M, Chin W, Garner E, Obradovic Z: **Intrinsically disordered proteins.** *J. Mol. Graph. Model* 2001, 19: 26-59
4. Demchenko AP: **Recognition between flexible protein molecules: induced and assisted folding.** *J. Mol. Recognit.* 2001, 14: 42-61
5. Uversky VN: **Natively unfolded proteins: a point where biology waits for physics.** *Protein Sci.* 2002, 11: 739-756
6. Tompa P: **Intrinsically unstructured proteins.** *TRENDS Biochem. Sci.* 2002, 27: 527-533
7. Fink AL: **Natively unfolded proteins.** *Curr. Opin. Struct. Biol.* 2005, 15: 35-41
8. Uversky VN, Oldfield CJ, Dunker AK: **Showing your ID: intrinsic disorder as an ID for recognition, regulation and cell signalling.** *J. Mol. Recognit.* 2005, 18: 343-384
9. Iakoucheva LM, Brown CJ, Lawson JD, Obradovic Z., Dunker AK: **Intrinsic disorder in cell-signalling and cancer-associated proteins.** *J. Mol. Biol.* 2002, 323: 573-584

10. Uversky VN, Oldfield CJ, Dunker AK: **Intrinsically disordered proteins in human diseases: introducing the D$^2$ concepts.** *Annu. Rev. Biophys.* 2008, 37: 215-246

11. Uversky VN: **Protein folding revisited. A polypeptide chain at the folding-misfolding-nonfolding cross-roads: which way to go?** *Cell and Mol. Life Sciences* 2003, 60: 1852-1871

12. Uversky VN, Fink AL: **Conformational constraints for amyloid fibrillation: the importance of being unfolded.** *Biochimica and Biophysica Acta* 2004, 1698: 131-153

13. Dunker AK, Cortese MS, Romero P, Iakoucheva LM, Uversky VN: **The roles of intrinsic disorder in protein interaction networks.** *FEBS Journal* 2005, 272: 5129-5148

14. Haynes C, Oldfield C, Ji F., Klitgord N, Cusick M, Radivojac P, Uversky VN, Vidal M, Iakoucheva L.: **Intrinsic disorder is a common feature of hub proteins from four eukaryotic interactomes.** *Plos Computational Biology* 2006, 2: 890-901

15. Dosztanyi Z., Chen J, Dunker AK, Simon I., Tompa P: **Disorder and sequence repeats in hub proteins and their implication for network evolution.** *Journal of proteome research* 2006, 5: 2985-2995

16. Kim PM, Sboner A, Xia Y, Gerstein M: **The role of disorder in interaction networks: a structural analysis.** *Molecular System Biology* 2008, 4: 179-186

17. Vucetic S, Obradovic Z, Vacic V, Radivojac P, Peng K, Iakoucheva L, Cortese M, Lawson J, Brown C, Sikes J, Newton C, Dunker AK: **DisProt: a database of protein disorder.** *Bioinformatics* 2005, 21: 137-140

18. Sickmeier M, Hamilton J, LeGall T, Vacic V, Cortese M, Tantos A, Szabo B, Tompa P, Chen J, Uversky VN, Obradovic Z, Dunker AK: **DisProt: the database of disordered proteins.** *Nucl. Acid Res.* 2007, 35: D786-D793

19. Vucetic S, Brown CJ, Dunker AK, Obradovic Z. **Flavors of protein disorder**. *Proteins* 2003, 52: 573-584

20. Ferron F, Longhi S, Canard B, Karlin D: **A practical overview of protein disorder prediction methods.** *Proteins* 2006, 65: 1-14

21. Dosztanyi Z, Sandor M, Tompa P, Simon I: **Prediction of protein disorder at the domain level.** *Curr. Prot. and Pept. Science* 2007, 8: 161-170

22. Daughdrill G, Pielak G., Uversky VN, Cortese M, Dunker AK: **Natively unfolded proteins.** In *Protein folding handbook*, J. Buchner and T. Kiefhaber, Weinheim, Wiley-VCH 2005: 275-337

23. Rose G. (Ed): **Unfolded proteins.** In *Advances in protein chemistry* 2002, 62: 1-398

24. Szilagyi A, Gyorffy D, Zavodszky P: **The twilight zone between protein order and disorder.** *Biophys. J.* 2008, 95: 1612-1626

25. Zhang Y, Stec B, Godzik A. **Between order and disorder in protein structures-analysis of "dual personality" fragments in proteins**. *Structure*. 2007, 15: 1141-1147

26. Uversky VN, Gillespie JR, Fink AL: **Why are "natively unfolded" proteins unstructured under physiological conditions?** *Proteins* 2000, 41: 415-427

27. Prilusky J, Felder C, Zeev-Ben-Mordehai T, Rydberg E, Man O, Beckmann J, Silman I, Sussman J: **FoldIndex: a simple tool to predict whether a given protein sequence is intrinsically unfolded.** *Bioinformatics* 2005, 21: 3435-3438

28. Garbuzynskiy SO, Lobanov M Yu, Galzitskaya OV: **To be folded or to be unfolded?** *Protein Sci.* 2004, 13: 2871-2877

29. Galzitskaya OV, Garbuzynskyi SO, Lobanov MY: **Prediction of natively unfolded regions in protein chain.** *Molecular Biology* 2006, 40: 298-304

30. Galzitskaya OV, Garbuzynskyi SO, Lobanov MY: **Prediction of amyloidogenic and disordered regions in protein chains.** *PLoS Comput. Biol.* 2006, 2: e177

31. Dosztanyi Z, Csimok V, Tompa P, Simon I: **The pairwise energy content estimated from amino acid composition discriminate between folded and intrinsically unstructured proteins.** *J. Mol. Biol.* 2005, 347: 627-639

32. Meszaros B, Tompa P, Simon I, Dosztanyi Z: **Molecular principles of the interactions of disordered proteins.** *J. Mol. Biol.* 2007, 372: 549-561

33. Shimizu K, Muraoka Y, Hirose S, Tomii K, Noguchi T: **Predicting mostly disordered proteins by using structure-unknown protein data.** *BMC Bioinformatics* 2007, 8: 78-92

34. Joachims T: **Transductive learning via Spectral Graph Transducer.** *Proceeding of International Conference on Machine Learning* 2003: 143-151

35. Obradovic Z, Peng K, Vucetic S, Radivojac P, Dunker AK: **Exploiting heterogeneous sequence properties improves prediction of protein disorder.** *Proteins* 2005, 7: 176-182

36. Peng K, Radivojac P, Vucetic S, Dunker AK, Obradovic Z: **Length-dependent prediction of protein intrinsic disorder.** *BMC Bioinformatics* 2006, 7: 208-225

37. Bordoli L, Kiefer F, Schwede T: **Assessment of disorder predictions in CASP7.** *Proteins* 2007, 69: 129-136

38. Ward JJ, Sodhi JS, McGuffin LJ, Buxton BF, Jones DJ: **Prediction and functional analysis of native disorder in proteins from the three kingdoms of life.** *J. Mol. Biol.* 2004, 337: 635-645

39. Oldfield C, Cheng Y, Cortese M, Brown C, Uversky VN, Dunker AK: **Comparing and combining predictors of mostly disordered proteins.** *Biochemistry* 2005, 44: 1989-2000

40. Bogatyreva NS, Finkelstein AV, Galzitskaya OV. **Trend of amino acid composition of proteins of different taxa**. J. Bioinform. Comput. Biol. 2006, 4: 597-608

41. Deiana A, Giansanti A: **Number of natively unfolded proteins scales with genome size.** *Biophysics and Bioengineering Letters* 2008, 1. http://padis2.uniroma1.it:81/ojs/index.php/CISB-BBL/article/view/2842/2918

42. Hobohm U, Scharf M, Schneider R, Sander C: **Selection of a representative set of structures from the Brookhaven Protein Data Bank.** *Protein Sci*. 1992, 1: 409-417

43. Hobohm U, Sander C: **Enlarged representative set of protein structure.** *Protein Sci*. 1994, 3: 522-524

44. Dosztanyi Z, Csizmok V, Tompa P, Simon I. **IUPred: web server for the prediction of intrinsically unstructured regions of proteins based on estimated energy content**. *Bioinformatics* 2005, 21: 3433-3434

45. Romero P, Obradovic Z, Xiaohong L, Gamer EC, Brown CJ, Dunker AK: **Sequence complexity of disorder proteins.** *Proteins* 2001, 42: 38-48

46. Higgs PG, Attwood TK.: *Bioinformatics and molecular evolution*. Blackwell Publishing 2006. Par. 10.2.

47. Smith DK, Radivojac P., Obradovic Z., Dunker AK, Zhu G. **Improved amino acid flexibility parameters.** *Protein Sci*. 2003, 12: 1060-1072

48. Baldi P, Brunak S, Chauvin Y, Andersen C, Nielsen H. **Assessing the accuracy of prediction algorithms for classification: an overview**. *Bioinformatics* 2000, 16: 412-424

49. Sing T, Sander O, Beerenwinkel N, Lengauer T. **ROCR: visualizing classifier performance in R**. *Bioinformatics* 2005, 21:3940-3941
50. Kumar S, Carugo O. **Consensus prediction of protein conformational disorder from amino acidic sequence**. *The Open Biochemistry Journal*. 2008, 2: 1-5
51. Lieutaud P, Canard B. Longhi S. **MeDor: a metaserver for predicting protein disorder**. *BMC Genomics*. 2008, 9: S25

# Tables

### Table 1 - Performance of single folding indexes on set A

Performance of *HQ*, mean packing $<P>$, mean pairwise energy $<E_c>$, *gVSL2* and Poodle-W in discriminating natively unfolded proteins from folded ones in the set A (Prilusky et al. [26]). Sensitivity ($S_n$), specificity ($S_p$), rate of false positives ($f_p$) and the area under curve (AUC) are defined in Methods. Mean packing threshold is fixed at 20.55 and mean pairwise energy threshold is fixed at -0.37 a.e.u.

|  | $S_n$ | $S_p$ | $f_p$ | AUC |
|---|---|---|---|---|
| *HQ* | 0.77 | 0.90 | 0.10 | 0.86 |
| $<P>$ | 0.82 | 0.87 | 0.13 | 0.90 |
| $<E_c>$ | 0.90 | 0.93 | 0.07 | 0.96 |
| *gVSL2* | 0.92 | 0.85 | 0.15 | 0.95 |
| Poodle-W | 0.74 | 0.93 | 0.07 | 0.94 |

### Table 2 - Performance of single folding indexes on set B

Performance of *HQ*, mean packing $<P>$, mean pairwise energy $<E_c>$, *gVSL2* and Poodle-W in discriminating natively unfolded proteins from folded ones among those in the set B (Shimizu et al. [32]). Mean packing threshold is set at 20.55 and mean pairwise energy threshold is set at -0.37 a.e.u.

|  | $S_n$ | $S_p$ | $f_p$ | AUC |
|---|---|---|---|---|
| Poodle-W | 0.74 | 0.98 | 0.02 | 0.95 |
| *gVSL2* | 0.81 | 0.94 | 0.06 | 0.94 |
| $<E_c>$ | 0.72 | 0.95 | 0.05 | 0.90 |
| $<P>$ | 0.70 | 0.90 | 0.10 | 0.85 |
| *HQ* | 0.70 | 0.92 | 0.08 | 0.86 |

### Table 3 - Performance of single folding indexes on set C

Performance of *HQ*, mean packing $<P>$, mean pairwise energy $<E_c>$, *gVSL2* and Poodle-W in discriminating natively unfolded proteins from folded ones among those in our own set C (see Methods). Mean packing threshold is fixed at 20.55 and mean pairwise energy threshold is fixed at -0.37 a.e.u.

|  | $S_n$ | $S_p$ | $f_p$ | AUC |
|---|---|---|---|---|
| Poodle-W | 0.75 | 0.87 | 0.13 | 0.90 |
| *gVSL2* | 0.83 | 0.76 | 0.24 | 0.86 |
| $<E_c>$ | 0.74 | 0.81 | 0.19 | 0.86 |
| <P> | 0.72 | 0.80 | 0.20 | 0.83 |
| HQ | 0.69 | 0.75 | 0.25 | 0.78 |

**Table 4 - Performance of each of the indexes combined into $S_{SU}$ on the sets B and C purged by the unclassified proteins**

Because of the strict unanimity clause the performances of the single indexes Poodle-W, gVSL2 and $<E_c>$, evaluated on a set purged by the unclassified proteins, coincides with that of $S_{SU}$, here reported on sets B and C. $nc$ is the fraction of unclassified proteins; $nc_f$ and $nc_u$ are the fraction of folded and unfolded proteins left unclassified.

| | $S_n$ | $S_p$ | $f_p$ | nc | $nc_f$ | $nc_u$ |
|---|---|---|---|---|---|---|
| Set B | 0.80 | 0.99 | 0.01 | 0.11 | 0.10 | 0.01 |
| Set C | 0.82 | 0.92 | 0.08 | 0.27 | 0.26 | 0.01 |

**Table 5 - Performance of strictly unanimous scores, as a function of the number of combined single indexes, tested on set C**

Performance of unanimous scores obtained by subsequently combining different indexes, starting from the one with the highest AUC, down to that with the lowest AUC. $S_{SU}^{(1)}$ refers to Poodle-W alone; $S_{SU}^{(2)}$ is obtained combining Poodle-W with gVSL2; $S_{SU}^{(3)}$ is the combination of Poodle-W, gVSL2 and $<E_c>$; $S_{SU}^{(4)}$ is the combination of Poodle-W, gVSL2, $<E_c>$ and $<P>$; finally $S_{SU}^{(5)}$ is the combination of Poodle-W, gVSL2, $<E_c>$, $<P>$ and HQ; $nc$ is the fraction of unclassified proteins.

| | $S_n$ | $S_p$ | $f_p$ | nc. |
|---|---|---|---|---|
| $S_{SU}^{(1)}$ | 0.75 | 0.87 | 0.13 | 0 |
| $S_{SU}^{(2)}$ | 0.84 | 0.88 | 0.12 | 0.18 |
| $S_{SU}^{(3)}$ | 0.82 | 0.92 | 0.08 | 0.27 |
| $S_{SU}^{(4)}$ | 0.81 | 0.93 | 0.07 | 0.31 |
| $S_{SU}^{(5)}$ | 0.84 | 0.93 | 0.07 | 0.39 |

# Figures

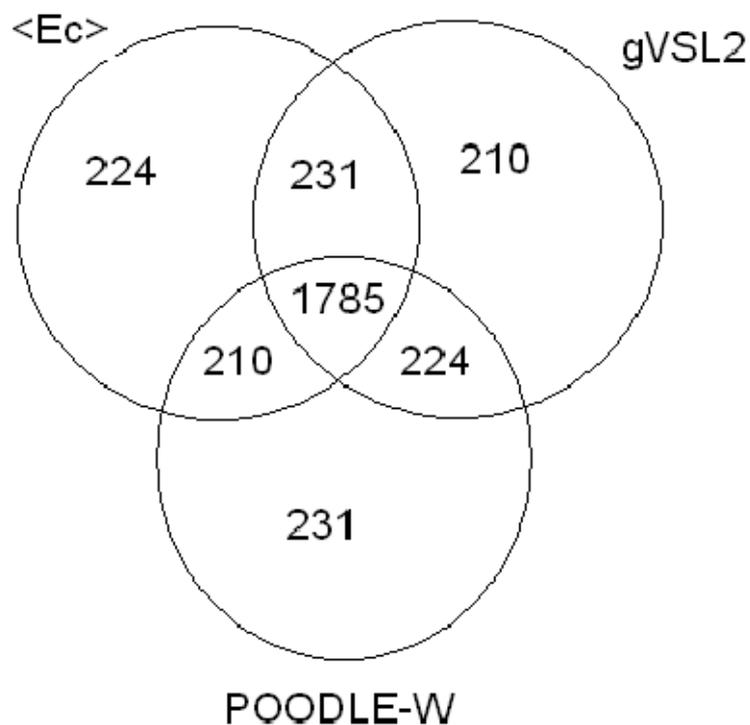

**Figure 1 - Overlaps between the predictions of single indexes**
Comparison of the predictions by mean pairwise energy, *gVSL2* and Poodle-W, on set C. The three indexes agree on 1785 proteins predicting them in the same class; 224, 210 and 231 proteins are singly classified by each of the indexes respectively, on each one of these proteins the prediction of one index is at variance with those of the others; the remaining figures refer to the pairwise cross-predictions.

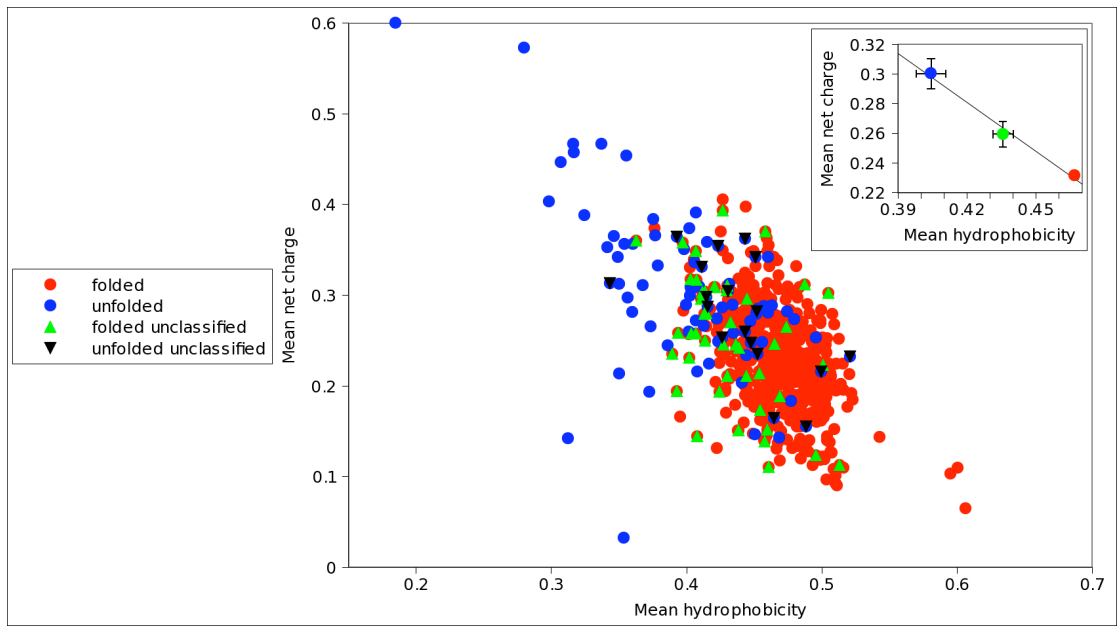

**Figure 2 - Hydrophobicity/charge plot of folded, unfolded and unclassified proteins in set B**

Hydrophobicity/charge plot of folded (red) and unfolded (blue) proteins in set B. Upper green triangles refer to folded unclassified proteins, lower black triangles refer to unfolded proteins unclassified by $S_{SU}$. This plot is a projection of the vector space of amino acidic compositions. Hydrophobicity and charge have been computed following ref. [25]. Folded proteins are reported in red, unfolded proteins in blue, folded proteins unclassified by $S_{SU}$ are reported in green, unfolded proteins unclassified by $S_{SU}$ are reported in black. Note that in this set, more *polarized* than set C, folded proteins have been chosen to have a minimum overlap with unfolded ones.

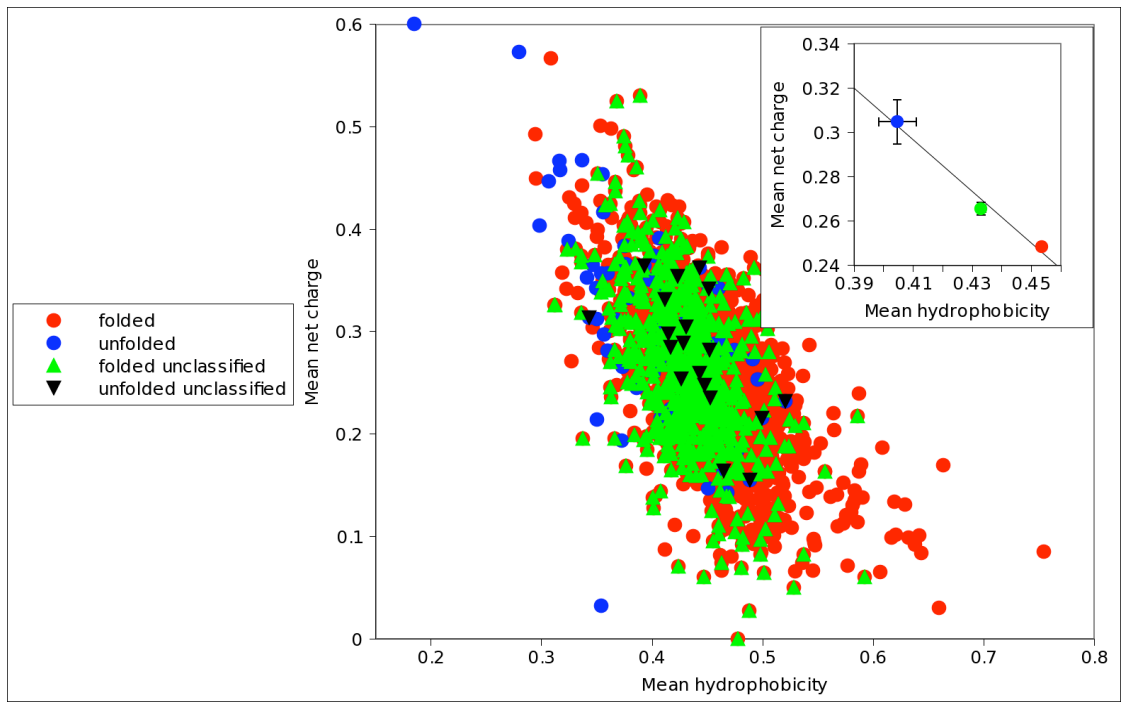

**Figure 3 - Hydrophobicity/charge plot of folded, unfolded and unclassified proteins in set C**

Hydrophobicity/charge plot of folded (red) and unfolded (blue) proteins in set C. Upper green triangles refer to folded unclassified proteins, lower black triangles refer to unfolded proteins unclassified by $S_{SU}$. Note the substantial overlap of folded (red) with unfolded proteins (blue), due to the lower polarization of this set with respect of set B (figure 2). Nevertheless, the centroids of the three distributions (see inset) are aligned and that of unclassified proteins is in between, indicating that the twilight zone is intermediate.

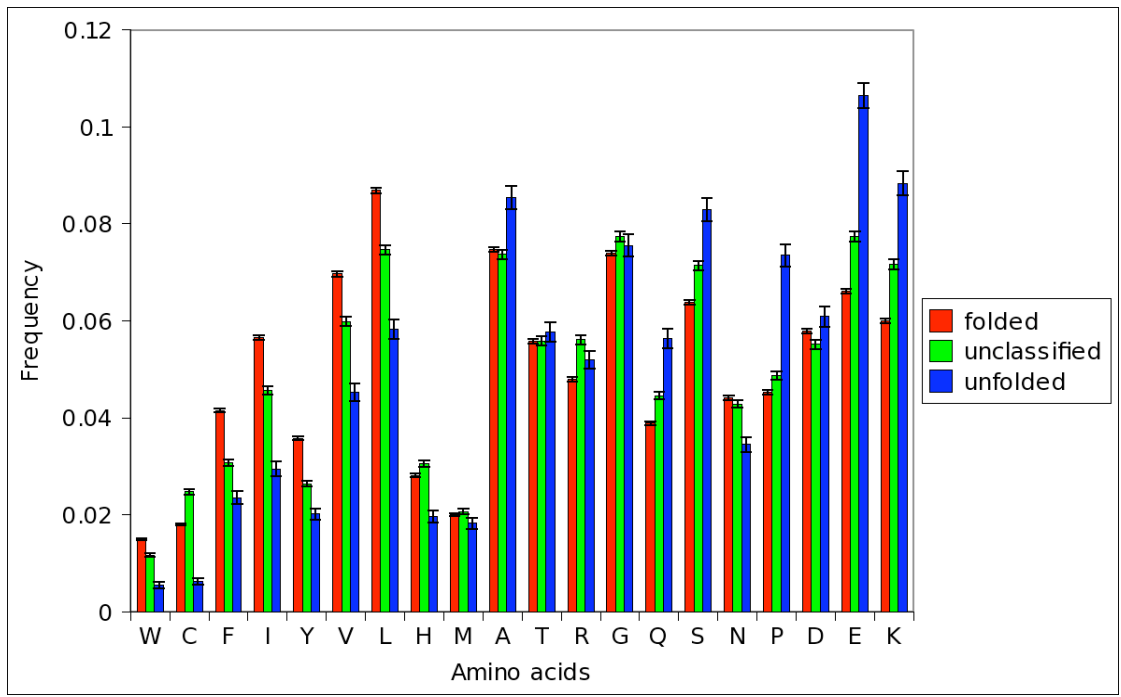

**Figure 4 - Amino acidic percent composition of folded, unfolded and unclassified proteins**

Amino acidic composition histograms of proteins in set C that are correctly predicted as: folded (red, left bars) and unfolded (blue, right bars) by $S_{SU}$; the center bars in green refer to unclassified proteins. The error bars are estimated from the variance formula given in the methods.

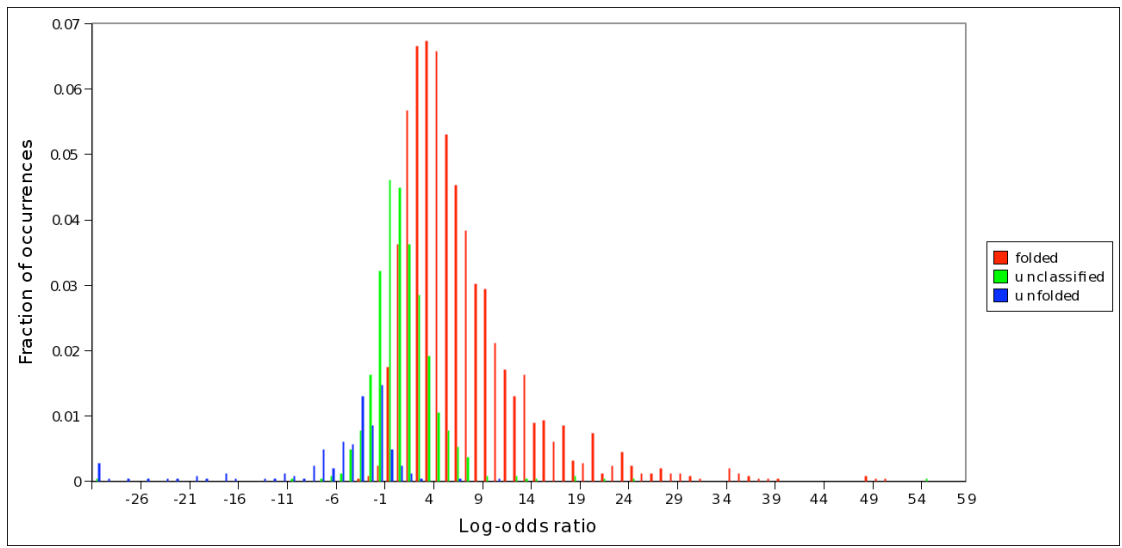

**Figure 5 - Distribution of the log-odds ratio S in folded, unfolded and unclassified proteins (twilight zone)**

Distribution of log-odds ratios in folded (red bars), unfolded (blue bars) and unclassified proteins (green bars), as evaluated by $S_{SU}$ on set C. From this graph the twilight zone can be defined as the set of proteins whose S-scores are sufficiently close to zero.

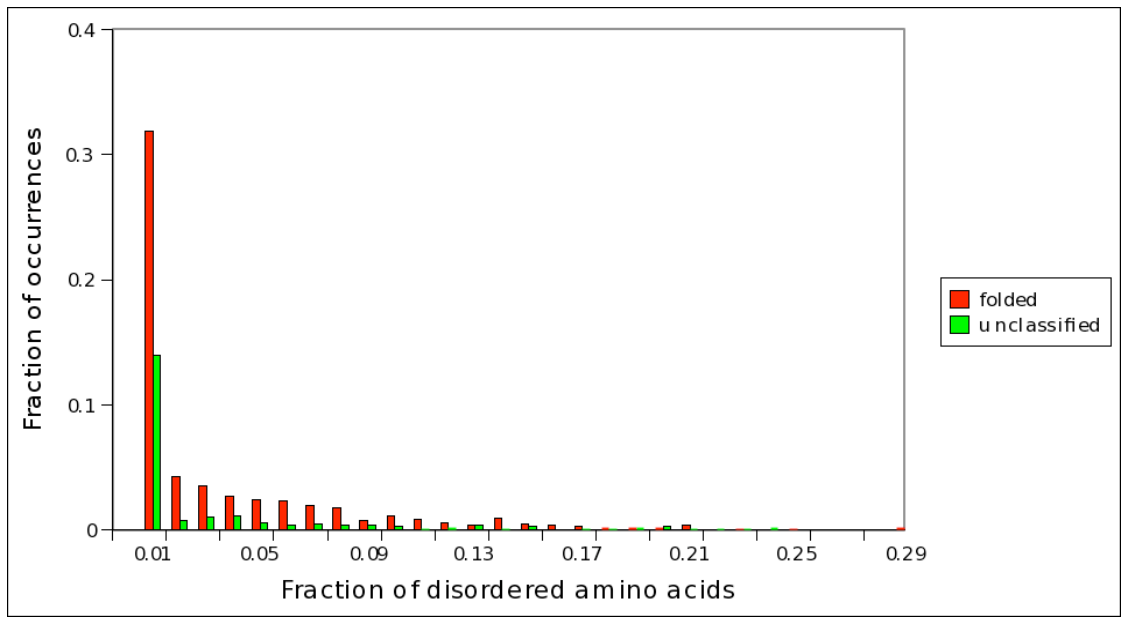

**Figure 6 - Fraction of disordered amino acids in folded, unfolded and unclassified proteins (twilight zone)**

Fraction of disordered amino acids in folded (red bars) and unclassified (green bars) proteins, as evaluated by $S_{SU}$ from set C. A residue is disordered if it is present in the SEQRES but not in the ATOM field in the PDB file of the protein [33].

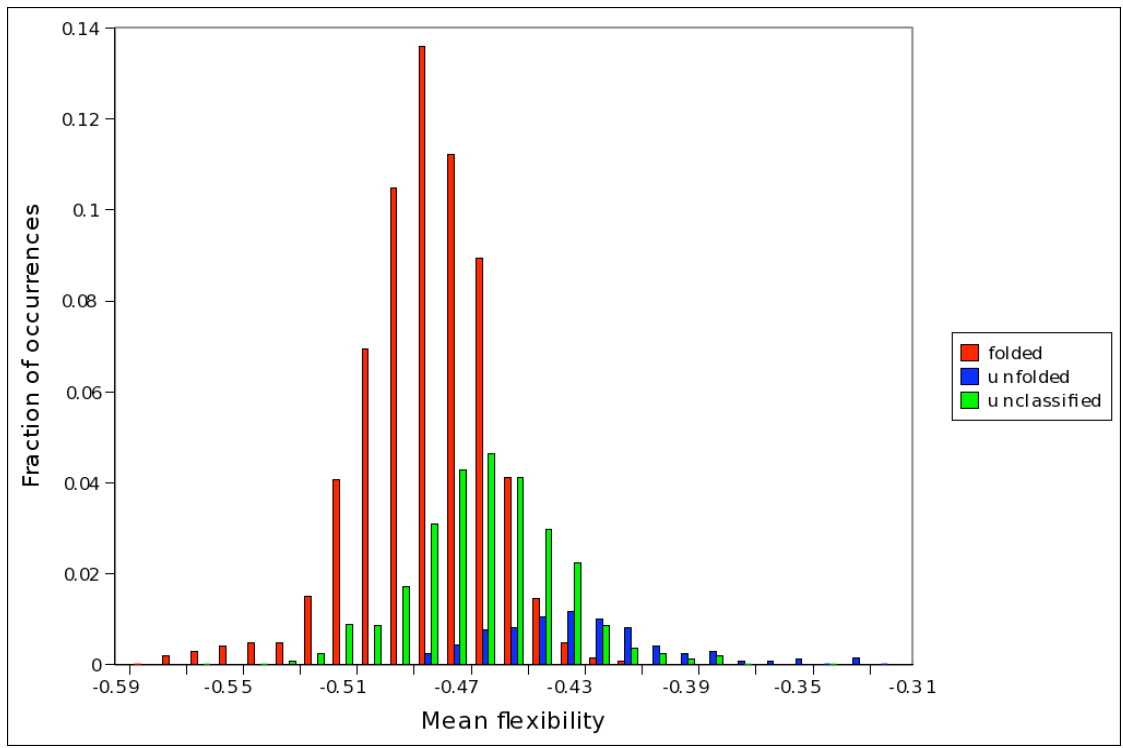

**Figure 7 - Distribution of mean flexibility in folded, unfolded and unclassified proteins (twilight zone)**

Distribution of mean flexibility in folded (red bars), unfolded (blue bars) and unclassified protein (green bars), as evaluated by $S_{SU}$ on set C.

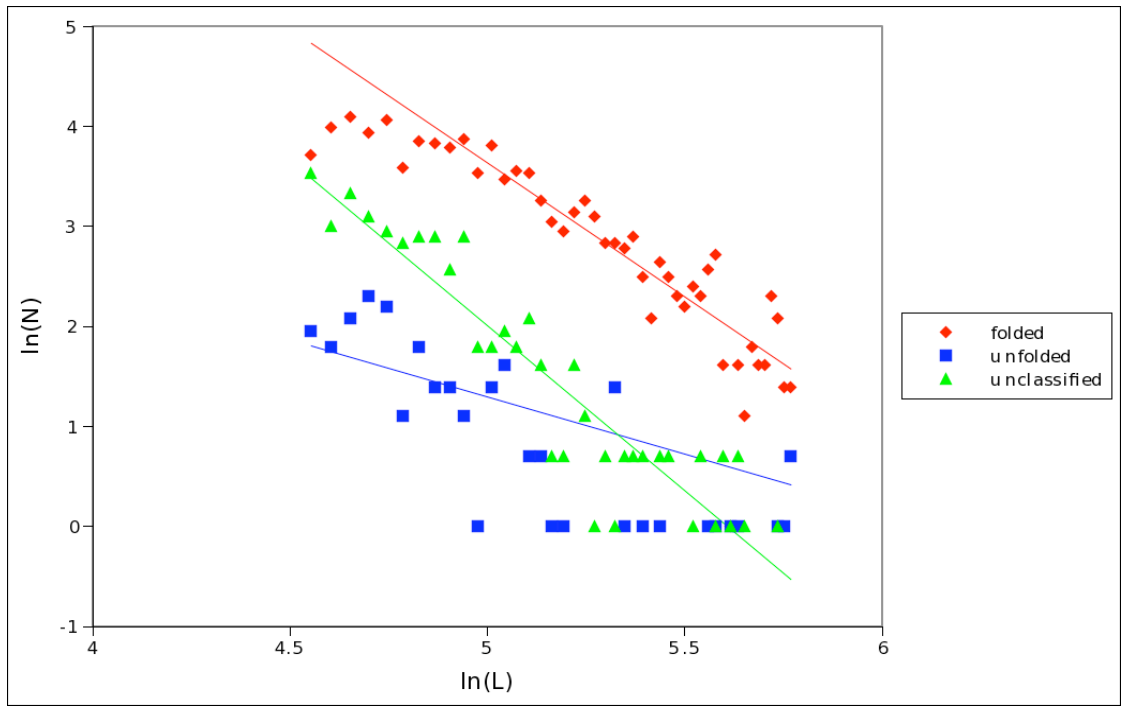

**Figure 8 - Distribution of lengths in folded, unfolded and unclassified proteins (twilight zone)**

Log-log plots of the distribution of lengths in the three classes of proteins extracted by $S_{SU}$ from set C. The scaling exponents, evaluated from a regression of the power law region in each graph are: -2.7 ± 0.2 (folded, red data points); -1.2 ± 0.3 (unfolded, blue); -3.3 ± 0.2 (unclassified, green).